\documentstyle[prl,aps,twocolumn]{revtex}
\begin{document}
\draft
%\twocolumn
\title{Universal Quantum Information Compression}

\author{Richard Jozsa \cite{poczta0}}
\address{School of Mathematics and Statistics\\
University of Plymouth, Plymouth, Devon PL4 8AA, England}

\author{Micha\l{} Horodecki \cite{poczta1}}

\address{Institute of Theoretical Physics and Astrophysics\\
University of Gda\'nsk, 80--952 Gda\'nsk, Poland}

\author{Pawe\l{} Horodecki \cite{poczta2}}
\address{Faculty of Applied Physics and Mathematics\\
Technical University of Gda\'nsk, 80--952 Gda\'nsk, Poland}

\author{Ryszard Horodecki\cite{poczta3}}
\address{Institute of Theoretical Physics and Astrophysics\\
University of Gda\'nsk, 80--952 Gda\'nsk, Poland}

\maketitle

\begin{abstract}
Suppose that a quantum  source is known to have von Neumann entropy
less than or equal to $S$ but is otherwise completely unspecified.
We describe a method of universal quantum data compression which
will faithfully compress the quantum information of any such source
to $S$ qubits per signal (in the limit of large block lengths).
%It
%follows that if the full knowledge of the density matrix of the
%source is unavailable then the Jaynes principle of maximum entropy
%characterises the optimal possible compression of the signals.
\end{abstract}

The question of compressibility of information is one of the
central issues in information theory. For classical information
Shannon's noiseless coding theorem \cite{Shannon,Cover} provides a
tight bound (equal to the Shannon entropy of the source) on the
extent to which information may be compressed. For quantum
information an analogous tight bound (equal to the von Neumann
entropy of the source) was established by Schumacher
\cite{Schumacher95} and further developed in \cite{JS,Barnum}. The
methods of information compression which are generally used to
establish these results, are {\em source specific} i.e. they apply
only to each given source separately. As elaborated below, the
classical compression protocol requires knowledge of the
probability distribution of the source and the quantum compression
protocol requires knowledge of the density matrix of the source. In
this letter we will consider the question of {\it universal}
quantum information compression. Is there a protocol which will
faithfully compress quantum information even if we do not know the
density matrix of the source? More precisely, suppose that all we
know about the source is that its von Neumann entropy does not
exceed some given value $S$. Is it then still possible to
faithfully compress the quantum information to $S$ qubits per
signal? Remarkably, in the case of classical information such
universal compression schemes are known to exist. An explicit
example is a scheme based on the theory of types developed by
Csiszar and K\"{o}rner \cite{Csiszar} (which is also described in
\S12.3 of \cite{Cover}). In this letter we will establish the
existence of universal compression schemes for {\it quantum}
information.

We begin with an outline of some source-specific compression
schemes which may be used to realise the Shannon and Schumacher
bounds. Later our main results will be related to an extension of
constructions occurring in these schemes. Consider a source of
classical information which generates signal $i$ with probability
$p_i$.
%MPRchange
Note that the signals may be faithfully represented using
$\log N$ bits/signal  by just using their names (here $N$ is the number
of signals; in this letter logarithms are always to base
2). Let $S= -\sum_i p_i \log p_i $ be the Shannon entropy of the
source (which is always $\leq \log N$). Shannon's theorem asserts
that the signals may be represented asymptotically faithfully using
only $S$ bits/signal and no fewer number of bits can suffice for
this task. Thus a sender (Alice) can communicate the sequence of
generated signals to a receiver (Bob) by sending $S$ bits/signal
and this transmission rate is optimal. The compression may be
achieved by the following method of {\it block coding} i.e.
processing long sequences of signals rather than individual signals
themselves separately. Note that we do not require that Bob is able
to recover the signals perfectly but only that the probability of
any error tends to zero in the limit of increasing block length.
(This is the meaning of the term ``asymptotically faithfully'').
Consider all possible signal sequences $i_1 i_2 \ldots i_n$ of
length $n$ (with associated probability $p_{i_1}p_{i_2} \ldots
p_{i_n}$). Let $SEQ(n)$ be the set of all such sequences of length
$n$. Our basic ingredient is the Theorem of Typical Sequences
\cite{Cover} which asserts the following:

\noindent
{\it Theorem of Typical Sequences:} For any given $\epsilon
>0$ and $\delta >0$ and for all sufficiently large $n$ there is a subset
$TY\!P(n) \subseteq SEQ(n)$ which has size $2^{n(S+\delta)}$ (i.e. an
exponentially small fraction of $SEQ(n)$) but whose total
probability exceeds $1-\epsilon$ (i.e. is as high as desired). The
sequences in $TY\!P(n)$ are called {\it typical} sequences and those
not in $TY\!P(n)$ are called {\it atypical} sequences.

 Intuitively
this theorem asserts that (for all sufficiently large $n$) any
sequence of signals generated by the source may be assumed with
arbitrarily high probability, to be a typical sequence. Thus to
achieve compression to $S$ bits/signal Alice and Bob set up a list
of names of all the typical sequences (requiring $n(S+\delta )$
bits per typical sequence).
%MPRchange
Then for sequences of length $n$ generated by the source Alice
sends the name of the sequence if it is a typical sequence and the
name of some fixed chosen typical sequence if it is atypical. In
the latter case Bob will be unable to regenerate the correct
message and an error will have occurred. However according to the
theorem of typical sequences, this can be arranged to occur with
arbitrarily small probability by choosing $n$ large enough.

The compression of quantum information was first considered by
Schumacher \cite{Schumacher95} who developed a quantum analogue of
Shannon's theorem. The quantum compression protocol was
subsequently simplified by Schumacher and Jozsa \cite{JS}
(hereafter referred to as the SJ protocol) and later Barnum et. al.
\cite{Barnum} showed that the limit of compression provided by the
SJ protocol is optimal i.e. that no other conceivable compression
protocol can provide further asymptotically faithful compression.

Consider a source of quantum states which produces pure states
$|\psi_i \rangle \in {\cal H}$ with probabilities $p_i$. Let $\rho
= \sum_i p_i |\psi_i \rangle \langle \psi_i |$ be the density matrix
of the source and let $S(\rho ) = - tr \rho \log \rho$ be its von
Neumann entropy. Then the SJ protocol \cite{JS} provides
asymptotically faithful compression to $S(\rho )$ qubits per signal
state. The method rests again on the theorem of typical sequences
above. Note that the density matrix of all signal sequences of
length $n$ is just $\rho^{\otimes n} = \rho \otimes \ldots \otimes
\rho$. Let $\lambda_i$ denote the eigenvalues of $\rho$ so that the
eigenvalues of $\rho^{\otimes n}$ are given by all products of the
form $\lambda_{i_1 \ldots i_n}= \lambda_{i_1} \ldots
\lambda_{i_n}$. Let $\Lambda (n)$ be the subspace of
${\cal H}^{\otimes n}$ given by the span of all eigenstates
$|\lambda_{i_1} \ldots \lambda_{i_n} \rangle$ corresponding to all
{\it typical} sequences $i_1 \ldots i_n$ of eigenvalues. $\Lambda
(n)$ is called the {\it typical subspace} (for block length $n$).
Since the Shannon entropy of the distribution $\lambda_i$ is equal
to the von Neumann entropy $S(\rho )$ we see that $\dim{} \Lambda (n)
= 2^{n(S(\rho )+\delta )}$ i.e. the typical subspace occupies about
$nS(\rho )$ qubits. Let $\Pi$ denote the projection onto the
typical subspace. Then by considering $\rho^{\otimes n}$ in its
eigenbasis and recalling the theorem of typical sequences we easily
see that
\begin{equation} \label{tr} tr \rho^{\otimes n} \Pi > 1-\epsilon
\end{equation}
This gives the SJ compression protocol: for sufficiently large $n$
Alice accumulates a sequence of $n$ signal states $|\psi_{in}
\rangle = |\psi_{j_1} \rangle \ldots |\psi_{j_n} \rangle $ and
performs a measurement which determines whether the joint state
lies in $\Lambda (n)$ or its orthogonal complement i.e. the joint
state is projected into one or other of these complementary
subspaces. If the state projects to $\Lambda (n)$ Alice sends the
resulting $n(S(\rho) + \delta )$ qubits to Bob. If it projects to
the orthogonal complement (which occurs with probability
$<\epsilon$) she sends to Bob any chosen state of $\Lambda (n)$.
%MPRchange
%If
%$\rho_{out}$ denotes the average state received by Bob then
%(\ref{tr}) implies that \[ \langle \psi_{in} | \rho_{out} |
%\psi_{in} \rangle > 1-\epsilon \] so that Bob receives the state
%$|\psi_{in} \rangle$ with arbitrarily high fidelity
%\cite{Schumacher95,Jozsa} and in the limit of $\delta
%\rightarrow 0$ only $S$ qubits/signal were transmitted.
%
Now, as proved in \cite{JS}, equation (\ref{tr}) implies that
 \[ \overline{ \langle \psi_{in} | \rho_{out} |
\psi_{in} \rangle }>1-2\epsilon \]
where $\rho_{out}$ is the state obtained by Bob if $|\psi_{in}
\rangle$ was generated by the source (and the average denoted by the
overbar is taken over all input blocks of signals $|\psi_{in}
\rangle $). Thus, Bob receives the state $|\psi_{in}
\rangle$ with arbitrarily high fidelity \cite{Schumacher95,Jozsa}
and in the limit of $\delta
\rightarrow 0$ only $S$ qubits/signal were transmitted.

We now come to the issue of {\it universal} compression. The above
compression schemes based on typical sequences and the typical
subspace are source-specific. For classical compression we need to
know the probability distribution of the source in order to
identify the typical sequences. For quantum compression we need to
know the density matrix of the source to identify the typical
subspace. Already here there is a significant difference: in the
quantum case we need to know neither the identity of the signal
states nor their probabilities, only their overall density matrix.
Our main result below will be more remarkable showing that we do
not even need to know the density matrix to achieve faithful
compression i.e. there exists a universal quantum compression
protocol which will faithfully asymptotically compress any quantum
source with von Neumann entropy $\leq S$ to $S$ qubits per signal.

But first, to illustrate the utility of the result, consider the
recently investigated problem of compression of quantum information
with incomplete data \cite{Jayncom}. Namely, suppose that the
information about the source is obtained via measurements performed
over a subensemble of the generated signal sequence. Suppose
further, that the set of measured observables was too small to
ensure a complete reconstruction of the density matrix of the
source. The question was: what is the maximal possible compression
rate $R$ allowing faithful transmission in this case? It has
been pointed out in \cite{Jayncom} that the Jaynes maximal entropy
principle \cite{Jaynes} places a lower bound on $R$:
\begin{equation}
R\geq S_J
\end{equation}
Here $S_J$ is maximal entropy admissible by the measured mean
values (Jaynes entropy). It has been also shown that for any {\it
qubit} source the
inequality passes into equality. Now,
%MPRchange
applying the universal quantum compression protocol we obtain that the
equality holds in the general case, so that the Jaynes entropy
gives the optimal compression with incomplete experimental data
characterizing the source.

We will now briefly outline a method of classical universal data
compression. Suppose we have a classical source and we know only
that its Shannon entropy is less than some given number $S$ (and we
do not know its probability distribution). Then a result of Csiszar
and K\"{o}rner \cite{Csiszar} shows that there exists a set of
sequences $CK(n) \subseteq SEQ(n)$ of length $n$ (whose description
depends only on the value of $S$) which satisfies all of the
properties enjoyed by $TY\!P(n)$ in the theorem of typical sequences
{\it not only for some one probability distribution with Shannon
entropy $S$ but simultaneously for all distributions with entropy
$\leq S$} i.e. the total probability of $CK(n)$ with respect to any
such distribution exceeds $1-\epsilon$ and the size of $CK(n)$ is
$2^{n(S+\delta )}$. The explicit construction of $CK(n)$ is also
described in \S 12.3 of \cite{Cover}. Hence if we replace $TY\!P(n)$
by $CK(n)$ in the classical compression scheme described previously
we will have a universal compression scheme which faithfully
asymptotically compresses any source with Shannon entropy $\leq S$
to $S$ bits/signal.

Consider next the prospect of replacing $TY\!P(n)$ by $CK(n)$ in the
SJ protocol. It is not difficult to see that this modified protocol
will faithfully compress to $S$ qubits/signal all those quantum
sources whose density matrices {\it commute with} $\rho$ and have
von Neumann entropy $\leq S$. Thus this does not provide a fully
universal quantum compression scheme: if we consider {\it all}
possible sources with von Neumann entropy $\leq S$ then
their density matrices need not commute. Below we describe an
alternative quantum compression scheme which is fully universal.

For any given $\rho$ let $\Xi$ be the subspace of ${\cal
H}^{\otimes n}$ in the modified SJ protocol, which is spanned by
all eigenstates of $\rho^{\otimes n}$ labelled by $CK$ sequences
i.e. $\Xi$ is the analogue of the typical subspace $ \Lambda (n)$.
Thus projection
%MPRchange
onto $\Xi$ will achieve faithful compression for
all sources with von Neumann entropy $\leq S$ whose density
matrices commute with $\rho$. A set of mutually commuting density
matrices is characterised by the corresponding common eigenbasis
and this may be any chosen orthonormal basis of $\cal H$. Thus as
$\rho$ varies over {\it all} possible density matrices with von
Neumann entropy $\leq S$ there will be a subspace $\Xi$ associated
with each choice of orthonormal basis of $\cal H$. We make this
dependence explicit by writing $\Xi (B)$ (where $B$ denotes an
orthonormal basis of $\cal H$) and we suppress explicit mention of
the values of $n$ and $S$ on which $\Xi$ also depends.

Now let $\Upsilon$ be the smallest subspace of ${\cal H}^{\otimes
n}$ which contains $\Xi (B)$ for {\it all} choices of basis $B$.
Then projection into $\Upsilon$ will achieve quantum compression
for all sources with von Neumann entropy $\leq S$. Below we will
prove that
\begin{equation} \label{dim} \dim{} \Upsilon \leq (n+1)^{d^2}
2^{n(S+\delta )} \end{equation} where $d= \dim{} {\cal H}$, $n$ is the
block length and $\delta >0$ may be as small as desired. Thus we
will achieve universal compression to $R$ qubits/signal where $R$
is given by \[ R= \lim_{n\rightarrow \infty} \frac{\log \dim{}
\Upsilon}{n} \leq \lim_{n\rightarrow \infty} d^2 \frac{\log (n+1)}{n}
+S+\delta \] which tends to $S+\delta$ qubits/signal. Since $\delta
$ can be as small as desired, asymptotically we have $S$
qubits/signal. This is our universal quantum information
compression scheme.

To prove (\ref{dim}) let $B^0 = \{ e^0 _1 , \ldots ,e^0 _d \}$ be
any fixed chosen orthonormal basis of $\cal H$. Then any other
basis $B= \{ e_1, \ldots ,e_d \}$ is obtained from $B^0$ by
applying some $d \times d$ unitary transformation $U$. Now $\Xi
(B)$ is the span of $2^{n(S+\delta )}$ states of the form $e_{i_1}
\otimes \ldots \otimes e_{i_n}$ (where we choose all $CK$ sequences
of the labels). Denote this basis by $CK(B)$. Hence $\Xi (B)$ is
precisely the subspace obtained by applying $U^{\otimes n}$ to $\Xi
(B^0 )$ (where $U^{\otimes n}$ is the unitary transformation on
${\cal H}^{\otimes n}$ given by $U\otimes \ldots \otimes U$).
%MPRchange
Then $\Upsilon$ is the  span of all $\Xi (B)$ as $B$ ranges over
all bases, which in turn equals the span of all $U^{\otimes n} \phi
$ where $U$ ranges over all $d\times d$ unitary matrices and $\phi$
ranges over $CK(B^0 )$.
Let
$M_d$ denote the linear space of all $d \times d$ complex matrices.
Since $M_d$ contains all unitary matrices
we get
\begin{equation} \label{updim}
\Upsilon   \subseteq   span \{ A^{\otimes n}\phi : A\in M_d, \, \phi
\in CK(B^0 ) \}
\end{equation}
For any fixed $\phi$ let
\begin{equation} \label{hphi} {\cal H}_\phi = span \{
A^{\otimes n} \phi : A\in M_d \} \end{equation} We will show that
\begin{equation} \label{hpdim} \dim{} \, {\cal H}_\phi \leq (n+1)^{d^2}
\end{equation}
Then using (\ref{updim}) and the fact that $\dim{} \, \Xi(B^0 ) =
2^{n(S+\delta )} $ we will immediately obtain our desired result
(\ref{dim}).

To prove (\ref{hpdim}) we use the notion of the symmetric subspace.

\noindent
{\it Definition.} The symmetric subspace of  a space ${\cal
H}^{\otimes n}$ is the space ${SY\!M}({\cal H})$ of the vectors
which are invariant under any permutation of the positions in the
tensor product.

The symmetric subspace has found various applications in quantum
information theory \cite{sym,Barenco}. In \cite{Barenco} it is
proved that the space $SY\!M({\cal H})$ has the following
properties:

(i) it is spanned by the vectors of the form $\psi^{\otimes n}$

(ii) its dimension is equal to $\left(n+d-1\atop d-1\right)$ where
$d=\dim{\cal H}$. In fact by considering the symmetrisation of a
product basis of ${\cal H}^{\otimes n}$ it is easy to obtain the
simpler over-estimate $\dim{} \, SY\!M({\cal H}) \leq (n+1)^d$
which will suffice for our purposes.

An important point to note is that for fixed $d$ and varying $n$
the size of $SY\!M({\cal H})$ grows only polynomially with $n$
whereas the full space ${\cal H}^{\otimes n}$ (of dimension $d^n$)
grows exponentially. Thus $SY\!M({\cal H})$ becomes exponentially
small inside ${\cal H}^{\otimes n}$ as $n$ grows.

Since $M_d$ is a linear space we can consider $M_d ^{\otimes n}$
and the symmetric subspace $SY\!M(M_d ) \subseteq M_d^{\otimes
n}$. According to (i) \[ SY\!M(M_d ) = span \{ A^{\otimes n} : A\in
M_d  \} \] and hence (\ref{hphi}) gives
\[ {\cal H}_\phi = span \{ B\phi : B\in SY\!M(M_d ) \} \]

Now, we can define  a linear mapping $\Gamma$ from the space
$SY\!M(M_d)$ to ${\cal H}_\phi$ by
\begin{equation}
SY\!M(M_d)\ni B\longrightarrow \Gamma (B) = B\phi\in {\cal H}_\phi
\end{equation}
This mapping is onto the space ${\cal H}_\phi$, and since it is
linear it cannot increase dimension. Hence  \[ \dim {\cal
H}_\phi\leq \dim SY\!M(M_d) \] Recalling that $\dim \, M_d = d^2$,
(ii) gives that
\[ \dim{} \, SY\!M(M_d) =\left(n+d^2-1\atop d^2-1\right) \leq (n+1)^{d^2} \]
which proves (\ref{hpdim}) and completes the proof of (\ref{dim}).

Thus we have shown that for any given $S$ and sufficiently large
$n$, projection into $\Upsilon (S,n)$ will provide universal
quantum data compression to $S$ qubits/signal for all sources of
pure quantum states with von Neumann entropy $\leq S$. The same
method will also work faithfully for all sources of {\it mixed}
states $\rho_i$ where the von Neumann entropy of $\rho =
\sum_i p_i \rho_i$ does not exceed $S$. Indeed we may always represent
each of these mixed states as a probabilistic mixture of pure
states whose identities we have forgotten. Also according to the
results of Barnum et. al. \cite{Barnum} our compression scheme is
optimal -- compression beyond $S$ qubits/signal cannot be faithful
for sources of entropy equal to $S$ and hence cannot be faithful
for all sources of entropy $\leq S$.

Finally we remark that our bound (\ref{dim}) on $\dim{} \Upsilon$,
although sufficient for our purposes, is not generally tight.
Indeed all we needed to show was that $\dim{} \Upsilon$ is some {\it
polynomial} (in $n$) multiple of $\dim{} \, \Xi (B^0 )$.
%MPRchange
It is interesting to note that
%the relation between the symmetric
%subspace and $\Upsilon$ in our scheme becomes tight
$\dim \Upsilon$ can be calculated exactly for the case of $S=0$.
Here we are considering all possible trivial sources $\Sigma
(\psi )$ which generate repeatedly one and the same vector $\psi$
(i.e. have von Neumann entropy zero). For $\Sigma (\psi )$ the
subspace $\Xi$ and the typical subspace are both just the one
dimensional $span \{
\psi \otimes \ldots \otimes \psi \} \in {\cal H}^{\otimes n}$. Hence
$\Upsilon$ is the span of all states of the form $\psi \otimes
\ldots \otimes \psi$ and by (i) we see that $\Upsilon$ equals
$SY\!M({\cal H})$ in this case. As noted previously, $SY\!M({\cal
H})$ becomes vanishingly small inside ${\cal H}^{\otimes n}$ as $n$
increases so the number of qubits per signal used for faithful
transmission tends to zero with increasing $n$.

M. H. and P. H. gratefully acknowledge the support from  Foundation
for Polish Science. R. J. is supported in part by the European TMR
network ERB-FMRX-CT96-0087.

\end{document}